# Dry Transfer Based on PMMA and Thermal Release Tape for Heterogeneous Integration of 2D-TMDC Layers

*Amir Ghiami,[†,*] Hleb Fiadziushkin,[†] Tianyishan Sun,[†] Songyao Tang,[†] Yibing Wang,[†] Eva Mayer,[‡] Jochen M. Schneider,[‡] Agata Piacentini,[§,α] Max C. Lemme,[§,α] Michael Heuken,[†,§] Holger Kalisch,[†] Andrei Vescan[†]*

[†] Compound Semiconductor Technology, RWTH Aachen University, 52074 Aachen, Germany

[‡] Materials Chemistry, RWTH Aachen University, 52074 Aachen, Germany

[§] Advanced Microelectronic Center Aachen (AMICA), AMO GmbH, 52074 Aachen, Germany

[α] Chair of Electronic Devices, RWTH Aachen University, 52074 Aachen, Germany

[§] AIXTRON SE, 52134 Herzogenrath, Germany

[*] Corresponding author: ghiami@cst.rwth-aachen.de

**ABSTRACT**: A reliable and scalable transfer of 2D-TMDCs (two-dimensional transition metal dichalcogenides) from the growth substrate to a target substrate with high reproducibility and yield is a crucial step for device integration. In this work, we have introduced a scalable dry-transfer approach for 2D-TMDCs grown by MOCVD (metal-organic chemical vapor deposition) on sapphire. Transfer to a silicon/silicon dioxide (Si/SiO$_2$) substrate is performed using PMMA (poly(methyl methacrylate)) and TRT (thermal release tape) as sacrificial layer and carrier, respectively. Our proposed method ensures a reproducible peel-off from the growth substrate and better preservation of the 2D-TMDC during PMMA removal in solvent, without compromising its adhesion to the target substrate. A comprehensive comparison between the dry method introduced in this work and a standard wet transfer based on potassium hydroxide (KOH) solution shows improvement in terms of cleanliness and structural integrity for dry-transferred layer, as evidenced by X-ray photoemission and Raman spectroscopy, respectively. Moreover, fabricated field-effect transistors (FETs) demonstrate improvements in subthreshold slope, maximum drain current and device-to-device variability. The dry-transfer method developed in this work enables large-area integration of 2D-TMDC layers into (opto)electronic components with high reproducibility, while better preserving the as-grown properties of the layers.

## 1. INTRODUCTION

Two-dimensional transition metal dichalcogenides (2D-TMDCs), recognized for their exceptional electronic, physical and chemical properties, are rapidly emerging on the forefront of research on solid-state micro-/nanoelectronic devices and circuits.[1] It has been suggested that these materials could extend Moore's law and pioneer innovative devices beyond CMOS (complementary metal-oxide-semiconductor) technology.[2] By 2028, the International Roadmap for Devices and Systems (IRDS) anticipates 2D-TMDC materials to find commercial applications in transistors and beyond-CMOS devices.[3]

Despite the impressive performance of devices based on 2D-TMDC layers created via mechanical exfoliation from bulk crystals, industrial scaling still presents a significant challenge. Synthesizing high-quality 2D-TMDCs requires elevated growth temperatures of 700-900 $^{o}C$, which exceed the CMOS back-end-of-line (BEOL) integration limit (< 400 $^{o}C$).[4] Additionally, optimized 2D-material growth is mostly performed on substrates different from conventional device templates like silicon/silicon dioxide.[5–9] This underscores the importance of developing robust and scalable methods to transfer 2D-TMDC films from their growth substrates onto device templates. Various methods have been developed to transfer 2D materials, which typically fall into "dry" and "wet" categories.[10–12] Within the community, there has been particular focus on dry methods in the last years, likely due to their scalability and automatability.[10] During the dry transfer of 2D-TMDC materials, a sacrificial layer is essential to provide a mechanical support and maintain the integrity of the 2D layer throughout the transfer process.[11] Polymers such as PMMA are commonly used for this purpose and are applied via spin-coating.[13–17] The mechanical properties of PMMA largely depend on its molecular weight (MW); therefore, high-MW PMMA (*e.g.* 950 K) is typically used to ensure sufficient mechanical support and prevent damage to the 2D material.[18,19] In the dry method, to peel-off the 2D-TMDC/PMMA stack from the growth substrate, a TRT (alternatively laser- or UV-release tape) serving as a carrier is attached on top of the spin-coated and baked polymer layer. The 2D-TMDC/polymer/tape stack is then delaminated through a mechanical peel-off.[12,16,20,21] Subsequently, the peeled stack is positioned onto the target substrate. The carrier tape is then delaminated by reducing its adhesion through heating (or laser/UV exposure).[12,16,22] Finally, the polymer is removed using chemical solvents.

While the dry-transfer steps may appear straightforward, this method still faces several obstacles, particularly in terms of yield and reproducibility. One issue is the process of peeling-off the 2D-TMDC/polymer stack from the growth substrate, which may result in the 2D-TMDC/polymer layer remaining partially or completely undetached. To address this, some methods incorporate a (semi)metallic interfacial layer (such as gold[23] or bismuth[24]) between the 2D-TMDC and the polymer. These layers, which adhere strongly to the 2D-TMDC and the polymer, are thought to improve peel-off yield. However, this adaption requires an additional metal etching step using acid at the end of the process, which can lead to further damage or contamination of the 2D-TMDC film.

Another complication arises in the final step during polymer removal with chemical solvents, during which the 2D-TMDC layer might partially or entirely detach from the target substrate which is rooted in a comparably weak adhesion.[10,20,25] This issue was addressed in a recent work by Kwon *et al.*,[20] in which a 2D-MoS$_2$ layer was successfully dry-transferred onto a prepatterned Si/SiO$_2$ substrate with Au contacts using PMMA and TRT. The high yield achieved after PMMA removal (in hot acetone) was facilitated by the strong adhesion between the Au surface and 2D-MoS$_2$. Additionally, they showed that the total thickness of the bottom contact should be maintained below 30 *nm* to preserve the 2D-MoS$_2$ during PMMA removal. However, their method is limited to prepatterned substrates only, and dry transfer of 2D-TMDCs on target substrates like Si/SiO$_2$ still remains a challenge due to the weak adhesion between SiO$_2$ and 2D-TMDC.

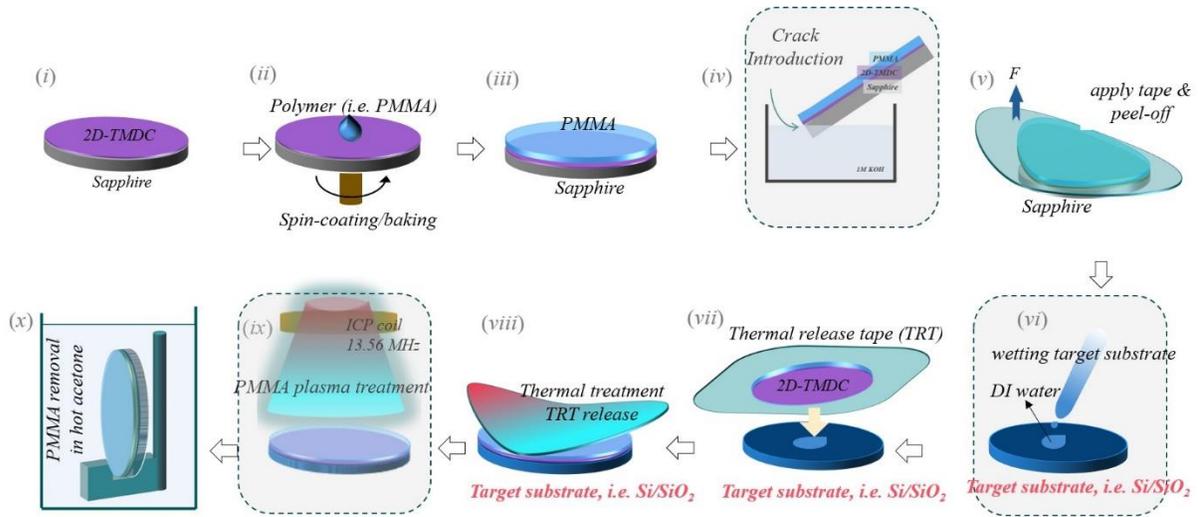

**Figure 1**. Detailed schematics of the dry transfer steps using PMMA and TRT; the steps developed in this study are outlined *(i)* as-received 2D-TMDC on growth substrate (*i.e.* sapphire), *(ii)* PMMA spin-coating and baking, *(iii)* PMMA spin-coated 2D-TMDC on sapphire, *(iv)* crack introduction at the edge of the stack between sapphire and 2D-TMDC/PMMA before applying the TRT, *(v)* TRT application and peel-off, *(vi)* wetting the target substrate with DI water droplet before transferring the peeled-off 2D-TMDC/PMMA/TRT stack. *(vii)* transfer the 2D-TMDC/PMMA/TRT stack onto the target substrate, *(viii)* TRT removal and annealing on hot plate, *(ix)* inductively coupled plasma (ICP) treatment of the PMMA layer, *(x)* PMMA removal in hot acetone.

Our work based on the dry-transfer approach using PMMA and TRT aims to tackle the described challenges by: (1) improving the peel-off of 2D-TMDC/polymer from the growth substrate (*i.e.* sapphire) and (2) a better preservation of the 2D-TMDC layer on the target substrate (*i.e.* Si/SiO$_2$) during the polymer removal with chemical solvents. The wafer-scale dry transfer process flow proposed in this work is schematically illustrated in Figure 1.

## 2. RESULTS AND DISCUSSION

2D-WSe$_2$ deposited on 2" sapphire substrates via a commercial AIXTRON CCS MOCVD (metal-organic chemical vapor deposition) tool was chosen as 2D-TMDC material in this work. For our studies, $3 \times 3$ cm$^2$ samples were cleaved from the wafer centers. Then, the PMMA was spin-coated on top and backed in air.

In the following step, in order to improve the peel-off process (the first aim of our work), we introduce a crack at the perimeter of the stack between growth substrate (sapphire) and the 2D-WSe$_2$/PMMA layer, using a KOH solution (in analogy with Fig. 1 *(iv)*). This was performed by $30\ s - 60\ s$ exposure to a 1 M KOH solution. The crack serves as the initiation point for the subsequent peel-off. As the KOH solution, unlike in wet-transfer methods,[17] only wets a small section of the 2D film edge, it does not impair the quality of the 2D-TMDC layer.

Next, a piece of TRT was applied onto the sample. Subsequently, a gentle mechanical peel-off was initiated by removing the 2D-WSe$_2$/PMMA/TRT, starting from the already cracked edge (cf. Fig. 1 *(v)*). Numerous experiments have demonstrated that such crack introduction at the edge of the stack consistently leads to a successful peel-off of the complete sample area. This is due to the fact that the majority of the force required is for the crack initiation (first stage of peel-off), which involves overcoming the inherent toughness of the material. Once initiated, crack propagation (second stage of peel-off) occurs with considerably less force, according to well-established principles in fracture mechanics.[21,26]

In the next step, inspired by direct wafer bonding techniques,[27,28] we wetted the Si/SiO$_2$ target substrate (with 100 $nm$ thermally grown SiO$_2$) with DI water droplets before transferring the peeled-off stack. DI water obviously displaces air and fills the gap between 2D-WSe$_2$ and the target substrate. It should be noted that, before transferring the stack, the extra tape overhang around the 2D-WSe$_2$ layer was cut away. The stack was then left for about 6 *h* at ambient temperature in a tilted position to allow the water to slowly diffuse laterally out of the interface, enabling the 2D-WSe$_2$ to conform to the target substrate and improve bonding.

After 6 *h* rest time, the stack was heated from room temperature to the 90$^o$C tape release temperature on a hot plate to remove the tape. Following tape removal, the stack was further heated to ~150$^o$C and maintained at that temperature for 10 *min* to further enhance the adhesion between 2D-TMDC and the target substrate (as illustrated in Fig. 1 *(viii)*).

In the following step, PMMA needs to be removed in hot acetone. In previous works, it was observed that performing this step directly causes partial detachment of the 2D-TMDC from the target substrate, particularly within the first minute of immersion. When PMMA is initially exposed to acetone, rapid solvent diffusion and polymer swelling lead to a significant release of mechanical stress in PMMA, [20,29] which, to our understanding, can disrupt the adhesion of the 2D material to the target substrate. By ensuring a slower, more controlled PMMA removal during this initial stage, this issue is expected to be resolved effectively. Therefore, the PMMA surface was treated with Ar/BCl$_3$ plasma in an inductively coupled plasma (ICP) chamber (as shown in Fig. 1 *(ix)*).[30] When PMMA is exposed to an inert plasma like Ar, high-energy ions and electrons can induce cross-linking on the polymer surface. This process, known as CASING (cross-linking by activated species of inert gas), involves the formation of covalent bonds between polymer chains, increasing the overall molecular weight of the surface layer.[30–32] In addition, BCl$_3$ plasma might further increase the molecular weight by forming a more cross-linked polymer. As intended, such a densified surface layer acts as a controlled barrier slowing down acetone penetration and successfully preventing TMDC detachment.

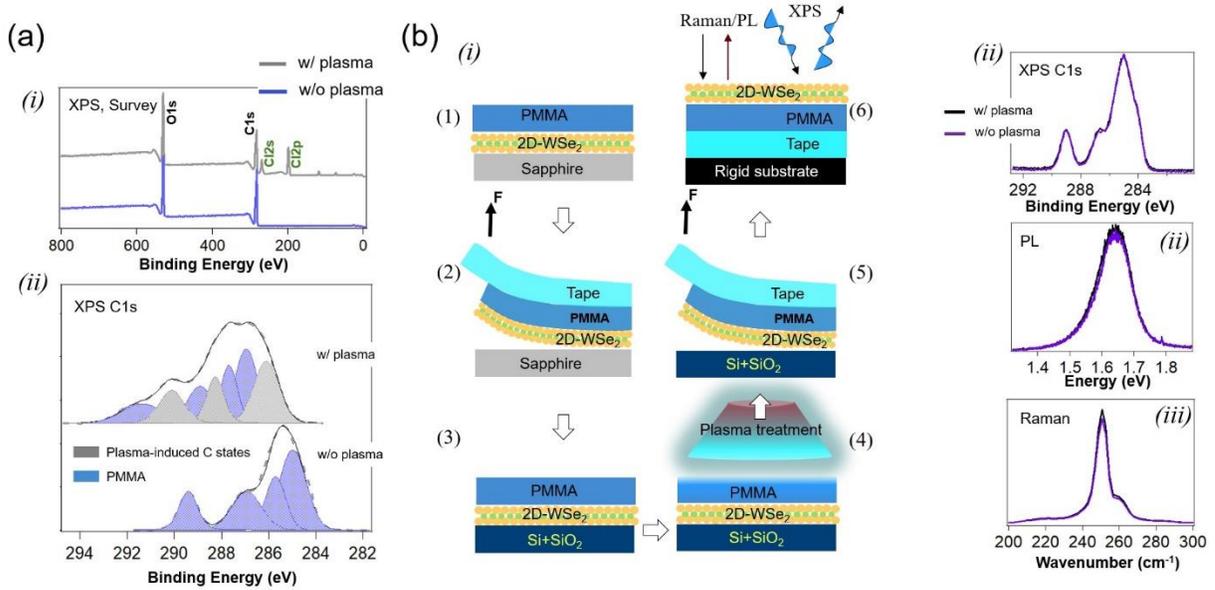

**Figure 2.** (a) Chemical analyses of plasma treated PMMA surface, *(i)* XPS survey scans, *(ii)* high resolution XPS C1s core level. (b) Exploring the potential impact of plasma treatment on the underlying 2D-WSe$_2$ and 2D-WSe$_2$/PMMA interface: *(i)* process steps to prepare 2D/PMMA/tape stack to access 2D and 2D/PMMA interface avoiding thick PMMA film, *(ii)* comparison of the XPS, PL and Raman signals of the 2D-WSe$_2$ and 2D-WSe$_2$/PMMA interface in plasma-treated and untreated stacks.

To verify our hypothesis of a densified PMMA surface and to exclude other effects of plasma treatment on the PMMA, X-ray photoemission spectroscopy (XPS) was performed. The XPS survey spectra of as-prepared PMMA and plasma-treated PMMA are shown in Figure. 2a *(i)*. In the as-prepared PMMA, only carbon and oxygen peaks are visible, while plasma-treated PMMA shows additional peaks attributed to chlorine (Cl), indicating the incorporation of Cl species during plasma exposure. The high-resolution *C1s* spectra of both PMMA samples are shown in Figure 2a *(ii)*. The as-prepared PMMA reveals characteristic PMMA peaks, highlighted in blue, with four distinct components: (1) hydrocarbon (C–C/C–H) at a binding energy of 285 eV, (2) β-shifted carbon at 285.7 eV, (3) methoxy group carbon at 286.8 eV and (4) carbon in the ester group at 289.1 eV.[33,34] The plasma-treated PMMA shows two key differences. First, the PMMA-associated peaks are shifted toward higher binding energies, which can be attributed to increased cross-linking and incorporation of a highly electronegative specie (*i.e.* Cl) caused by plasma-induced modifications.[30] Additionally, three new peaks appear, likely due to plasma-induced carbon states associated with bond formations between carbon and Cl or cross-linked carbon structures, consistent with the changes expected from Ar/BCl$_3$ plasma treatment.[35–37]

To further confirm that plasma treatment only affects the surface of the PMMA without altering the 2D-TMDC layer or the 2D-TMDC/PMMA interface, we conducted XPS, Raman and photoluminescence (PL) spectroscopy (Fig. 2b). For this purpose, two sets of 2D-WSe$_2$/PMMA stacks were prepared: one with PMMA plasma treatment and one without. The stack preparation steps are shown in Figure. 2 b *(i)* and include a second transfer, exposing the 2D-WSe$_2$ film on top of PMMA (either treated or untreated)/TRT. This approach allows accessing the 2D-WSe$_2$ and the 2D-WSe$_2$/PMMA interface directly from the TMDC side. The XPS, PL and Raman spectra of both stacks show negligible differences, suggesting an unaffected 2D-WSe$_2$ and 2D-WSe$_2$/PMMA interface after plasma treatment. (Fig. 2b *(ii)*)

Figure 3a illustrates the main transfer steps for a 3 × 3 cm$^2$ 2D-WSe$_2$/sapphire sample. Figure 3a *(iv)* shows the successfully transferred 2D layer on SiO$_2$/Si after PMMA removal, with no visible macroscopic defects or pinholes as evaluated by optical microscopy (Fig. 3b *(iv)*). The improved adhesion of the 2D-TMDC to Si/SiO$_2$, along with the more controlled PMMA removal –achieved through wetting the target substrate and PMMA plasma treatment– contributed to the preservation of the 2D-TMDC. Without these measures, partial detachment of the 2D-TMDC layer frequently occurs; examples of such failures are presented in Figure 3b *(i)-(iii)*. Figure 3c *(i)* and *(iii)* display the Raman and PL maps of the transferred 2D-WSe$_2$ layer, respectively, along with the corresponding spectra (Fig. 3c *(ii)* and *(iv)*). Both the Raman and PL data confirm the uniformity and superior integrity of the transferred layer.

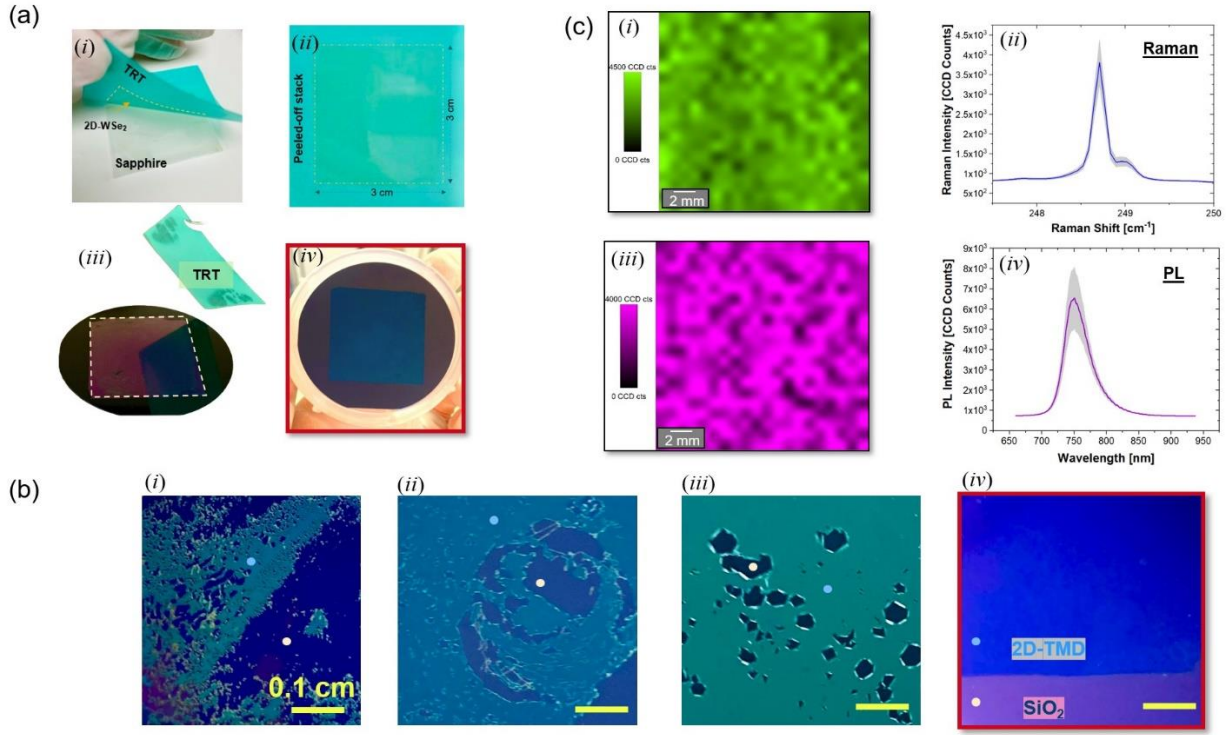

**Figure 3**. (a) Main dry transfer steps for a 3 × 3 cm$^2$ 2D-WSe$_2$/sapphire sample: *(i)* peel-off, *(ii)* peeled 2D-WSe$_2$/PMMA/TRT stack, *(iii)* removing the tape and annealing the stack on a hot plate, *(iv)* 2D-WSe$_2$ layer on target substrate (Si/SiO$_2$) after PMMA removal. (b) Impact of target substrate wetting and PMMA plasma treatment on the dry transfer yield: The optical microscope images show the 2D-TMDC after transfer and PMMA removal in hot acetone; *(i)* without substrate wetting and plasma treatment, *(ii)* with substrate wetting but no plasma treatment, *(iii)* with plasma treatment but no substrate wetting; in all three cases, sever 2D-TMDC detachment and pinholes are visible. *(iv)* with both substrate wetting and plasma treatment, achieving 100% transfer yield with no detachment/pinholes visible under optical microscope (image b *(iv)* highlights an area near the edge of the transferred layer to better visualize the 2D-TMDC and substrate). In (b), blue and pink spots point on 2D-TMDC and target substrate regions, respectively. (c) Raman and PL characterization of the dry-transferred 2D-WSe$_2$ layer: *(i)* Raman map and *(ii)* corresponding spectrum with raw data in greyscale; *(iii)* PL map and *(iv)* corresponding spectrum with raw data in greyscale.

Additionally, to further demonstrate the efficacy of the developed dry transfer method we transferred a 2D-MoS$_2$ layer onto a prepatterned substrate with 50 *nm* Au contacts, comparable to Kwon *et al.*[20]. Even on a structured template, our transferred layer showed near-perfect transfer yield with no defects or pinholes visible under optical microscope after PMMA removal (see Supporting Information).

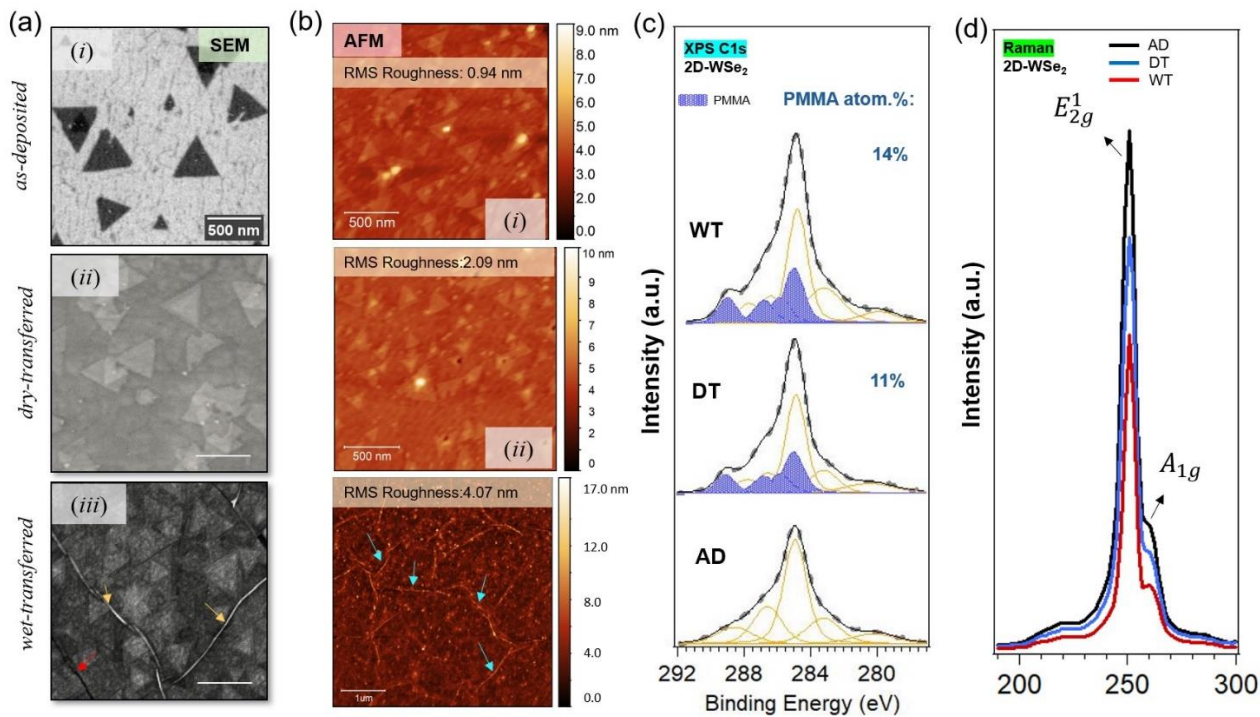

**Figure 2.** Characterization of the as-deposited and transferred 2D-WSe$_2$ layers. (a) SEM images of *(i)* as-deposited, *(ii)* dry-transferred and *(iii)* wet-transferred 2D-WSe$_2$. In the wet transferred sample, wrinkles and cracks are highlighted by yellow and red arrows, respectively. The scale bar size is 500 nm for all images. (b) AFM images of *(i)* as-deposited, *(ii)* dry-transferred and *(iii)* wet-transferred 2D-WSe$_2$. In the wet transferred layer, the 2D-TMDC nuclei are not visible due to a high density of wrinkles. (c) High resolution XPS *C1s* spectra of as-deposited versus dry- and wet-transferred 2D-WSe$_2$ (after PMMA removal). (d) Raman spectra of as-deposited versus dry- and wet-transferred 2D-WSe$_2$. As-deposited, dry-transferred and wet-transferred samples are denoted as AD, DT and WT, respectively.

In order to assess the quality and cleanliness of the 2D-WSe$_2$ layers dry-transferred onto Si/SiO$_2$ substrate and to provide a comparison with those after standard KOH-assisted wet transfer (described by Schneider *et al.*[38]), scanning electron microscopy (SEM), atomic force microscopy (AFM), Raman spectroscopy, and XPS were performed. Figure 4a *(i)* presents an SEM image of the as-deposited 2D-WSe$_2$, which is a coalesced monolayer with approx. 30% bilayer coverage. Figure 4a *(ii)* and *(iii)* show the dry- and wet-transferred layers after PMMA removal. The dry-transferred layer appears free of wrinkling and cracking. In contrast, the wet-transferred counterpart (Fig. 4a *(iii)*) exhibits wrinkles and cracks formed during wet transfer (Fig. 4a *(iii)*).

AFM images of the as-deposited, dry- and wet-transferred layers are presented in Figure 4b *(i)*-*(iii)*, respectively. A root mean square (rms) roughness of 0.94 nm was measured for the as-deposited layer. The dry-transferred layer exhibits an increase in rms roughness to about 2 nm, likely due to some PMMA residues remaining on the surface. The wet-transferred layer shows an even higher rms roughness of approximately 4 nm, probably to be attributed to the higher amount of polymer residues and the presence of wrinkles (indicated by blue arrows).

XPS analyses of the as-deposited and transferred layers have been performed. The fitted high-resolution XPS *C1s* core levels of the as-deposited as well as wet- and dry-transferred layers are shown in Figure 4c. The detected carbon signals can be referred to three different sources, (1) co-deposited carbon during MOCVD,[39] (2) adventitious carbon and (3) PMMA. The peaks attributed to the first two sources are shown in yellow. The PMMA-associated peaks filled in blue are resolved in four components as descried previously. The amount of PMMA residues calculated based on its atomic percentage are also shown in Figure 4c. It was observed that dry-transferred layers contained up to 23% less PMMA relative to wet-transferred ones. Although dry transfer reduces PMMA residues to some extent, a considerable amount still remains, which can adversely affect the performance of devices based on these transferred layers.[40,41] Addressing this challenge is critical for future improving 2D-TMDC-based devices. More details on the chemical composition of the as-deposited, dry- and wet-transferred 2D-WSe$_2$ layers are provided in the Supporting Information.

Raman spectra of the as-deposited, dry-transferred, and wet-transferred layers are depicted in Figure 4d. The characteristic WSe$_2$ Raman peaks, namely $E^1_{2g}$ and $A_{1g}$, were observed at 250 *cm$^{-1}$* and 261 *cm$^{-1}$*, respectecely.[42,43] The spectra revealed a lower intensity for wet-transferred layer in comparison to the dry-transferred counterpart. This reduction in intensity could be due to three factors: (1) PMMA residues may partially absorb or scatter incoming and reflected light; (2)

molecular interactions between PMMA and the underlying 2D film; and (3) a better preservation of layer quality and reduced damage after the dry transfer method.[25,44]

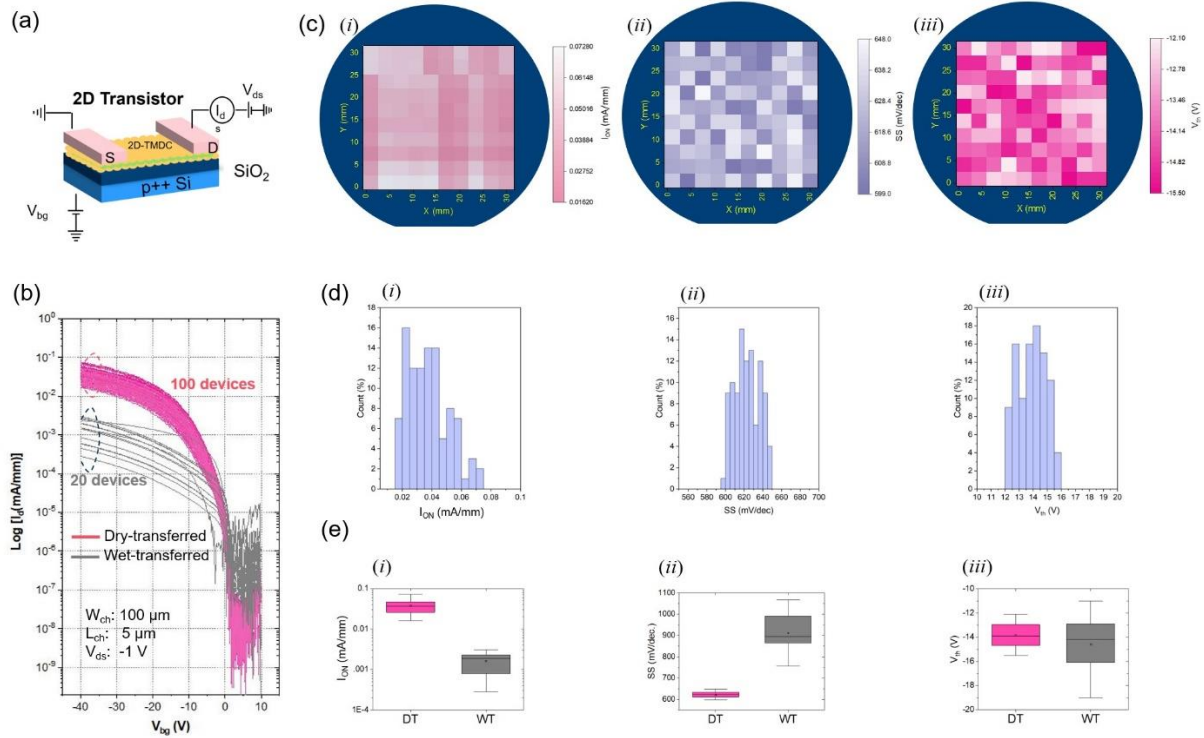

**Figure 3.** Field-effect transistors based on dry- and wet-transferred 2D-WSe$_2$ layers: (a) schematic illustration of globally back-gated field-effect transistors used in this study, (b) *I-V* transfer characteristics of transistors based on dry- and wet-transferred 2D-WSe$_2$ layers. (c) Spatial map of *(i)* ON current, *(ii)* subthreshold slope, and *(iii)* threshold voltage variations within the 3 × 3 cm$^2$ region of transistors fabricated using dry-transferred layer. (d) Histograms showing the distribution of *(i)* ON current, *(ii)* subthreshold swing, and *(iii)* threshold voltage values of the maps in (c). (e) comparison of the device metrics for transistors based on dry- and wet-transferred layers. Dry-transferred and wet-transferred are denoted as DT and WT, respectively.

To evaluate and compare the electrical performance of 2D-WSe$_2$ layers obtained via dry and wet transfer, field-effect transistors (FETs) were fabricated. A large 3 × 3 cm$^2$ dry-transferred layer (shown in Fig. 3a *(iv)*) and a 1 × 1 cm$^2$ wet-transferred layer were used for device fabrication and characterization. Gold (Au) was used as the top contact, with highly p-type doped Si serving as a global back gate with 100 nm thermally-grown SiO$_2$ layer as the gate dielectric. The devices were measured under ambient condition. The FET layout is schematically illustrated in Figure 5a (channel length ($L_{Ch}$) of 5 μm, channel width ($W_{Ch}$) of 100 *μm)*.

To assess device metrics, ON currents ($I_{ON}$) were extracted at a drain bias $V_{DS}= -1$ V, and back-gate voltage $V_{BG}= -40$ V.[45] Moreover, the subthreshold slope (*SS*) was determined based on the following equation:[46,47]

$$SS = \frac{d\,V_{BG}}{d\,log I_D} \quad (1)$$

, where $I_D$ is the drain current. In addition, the threshold voltage ($V_{th}$) of the devices was extracted using the linear extrapolation method.[48]

In total, 100 FETs across a 3 × 3 cm² area were measured for the dry-transferred layer, and 20 devices based on wet-transferred layer (1 × 1 cm²) were evaluated. The *I-V* transfer characteristics of all devices are shown in Figure 5b. Figure 5c *(i)-(iii)* displays the spatial variation maps for $I_{ON}$, *SS* and $V_{th}$ of the FETs based on dry-transferred 2D-WSe$_2$. The corresponding distributions are shown in the histograms in Figure 5d *(i)-(iii)*. The variations of device metrics indicate that the device performance is uniform over the 3 × 3 cm² area.

The comparison of device metrics, namely $I_{ON}$, *SS*, and $V_{th}$, between wet- and dry- transferred layers is summarized in Figure 5e *(i)-(iii)*. The FETs based on dry-transferred layers demonstrate significant improvement in the extracted device parameters, with an over 50× increase in the average $I_{ON}$ compared to wet-transferred counterpart. (Fig. 5e *(i)*) For the SS, devices based on dry-transferred layers exhibit values approximately 1.5× smaller than those based on wet-transferred layer. Additionally, the threshold voltage for devices based on dry transferred layers shows a significantly narrower distribution.

Device-to-device variability for the device metrics was also quantified using the coefficient of variation (*CV*), calculated as the ratio of the standard deviation to the mean value ($CV = \frac{\sigma}{\mu}$). The *CV* for $I_{ON}$ of dry- and wet-transferred was found to be 0.338 and 1.124, respectively. For the SS, the *CV* values for dry- and wet-transferred were 0.2428 and 0.3053, respectively. As for threshold voltage, the obtained CV values were found to be 0.4852 and 0.6247, for devices based on dry- and wet- transferred layers, respectively. These results demonstrate that dry-transferred layers provide more consistent performance with lower device-to-device variability. The improvement in electrical performance observed for dry-transferred layers can be attributed to the absence of wrinkles, cracks and KOH-related degradation,[17] and a lower degree of polymer contamination.

## 3. CONCLUSION

In this work, we have developed a dry transfer method for 2D-TMDC materials utilizing widely available materials, PMMA and thermal release tape. This method effectively addresses the challenges associated with preserving 2D-TMDC layer during PMMA removal, ensuring minimal damage and maintaining layer integrity. Compared to the standard KOH-assisted wet transfer process, the dry method better retains the quality of the as-deposited 2D-TMDC layer, as confirmed by Raman, XPS, AFM and SEM characterization. This preservation translates into improved transistor performance, including enhanced ON currents, lower subthreshold slopes, narrower threshold voltage distributions, and a reduced device-to-device variability. Despite these advancements, PMMA residues still remain a significant challenge, impacting the cleanliness of the transferred layers and the overall reliability of 2D-TMDC-based devices. Future efforts should focus on identifying alternative polymers with high solubility to achieve cleaner interfaces while maintaining sufficient mechanical stiffness to prevent 2D-TMDC damage during transfer.

## 4. EXPERIMENTAL SECTION

*MOCVD of 2D-WSe$_2$*: The 2D-WSe$_2$ layer was epitaxially grown on sapphire (0001) using a commercial AIXTRON Close-Coupled Showerhead (CCS) reactor in 7×2" configuration. The sapphire was first desorbed at $1050^oC$ in a 150 *hPa* H$_2$ atmosphere. Tungsten hexacarbonyl (W(CO)$_6$) and di-*iso*-propyl selenide (DiPSe) were used as tungsten and selenium precursors, respectively, with $H_2$ as the carrier gas at 20 *hPa* reactor pressure. The growth temperature was 720°C, and the flow rates of W and Se precursors were 25 nm/min and 110 µmol/min, respectively.

*Raman and photoluminescence spectroscopy*: Non-resonant Raman measurements were performed at room temperature using a WiTec confocal microscope with a solid-state 532 nm laser. The laser line was focused on the sample by a 100× microscope objective lens. The collected light was dispersed by a grating with 1800 grooves/mm. The laser power was set to 1*mW* to ensure a high signal-to-noise ratio and prevent laser-induced heating. The PL data was acquired with the same system using a 300 grooves/mm grating.

*X-ray photoemission spectroscopy*: For the XPS measurements, an AXIS Supra instrument (Kratos Analytical Ltd.) was used. Monochromatic X-rays are generated by an *Al Kα* source with an excitation energy of 1,486.6 eV. During data acquisition, charge neutralization with an electron-only source was performed to compensate for any charging effect. The high-resolution spectra were acquired with pass energy 10 eV and a step size of 0.05 eV. For quantitative analyses, the spectra were acquired with 20 eV pass energy and 0.1 eV step size. The binding energies were calibrated with respect to the *C1s* core level of hydrocarbons associated to adventitious carbon located at 284.8 eV.[49] For XPS core level analyses, a Shirley function for background subtraction followed by Voigt lineshape deconvolution was used.[39,50]

*Dry transfer of 2D-TMDC layers*: In dry transfer, *950 PMMA A6*, obtained from *Kayaku Advanced Materials, Inc*, was spin-coated at 3,000 *rpm* for 60 *s* on an as-grown 2D-TMDC/sapphire sample, and baked on a hot plate at 120 °*C* for 10 *min* in air (Figure 1c *(ii)*). As for the peel-off, the TRT applied was supplied by Nitto Revalpha (*no. P/N 319Y-4LS*) with an adhesion force of 4.8 N/20 mm on polyethylene terephthalate (PET). The PMMA plasma treatment (Fig. 1 (*ix*)) was performed in an ICP chamber (*SENTECH SI 500*) with a $BCl_3$/Ar plasma for 1 min, using gas flows of 10 and 20 standard cubic centimeters per minutes (*sccm*), for BCl$_3$ and Ar, respectively (ICP and RF power: 30 W).

*Device fabrication*: For the fabrication of globally back-gated 2D-WSe$_2$ field-effect transistors, after transferring (either wet or dry) the 2D layers onto p-type Si with 100 *nm* of thermally grown SiO$_2$ and removing the PMMA, metal contacts were defined by photolithography, metal deposition (Au, 80 *nm*), and lift-off processes. To define the device region, a photoresist layer was patterned by photolithography. Then, BCl$_3$ plasma (ICP and RF power: 30 W, 20 *sccm* BCl$_3$ + 10 *sccm* Ar) was used to etch away the unprotected 2D-TMDC layer before the photoresist layer was removed. More details of the device fabrication protocol are explained elsewhere.[25]

ASSOCIATED CONTENT

**Supporting Information**

Detailed description of the in-depth XPS analysis of 2D layers and the dry transfer onto prepatterned substrate are provided in the supporting information.

AUTHOR INFORMATION

**Corresponding Author**

Amir Ghiami – *Compound Semiconductor Technology, RWTH Aachen University, 52074 Aachen, Germany*

**Author Contributions**

A.G. conceived the project and designed the experiments. A.G. carried out experiments and data analysis. H.F. helped with the SEM measurement and prepatterned substrate fabrication. T.S. helped with the electrical measurements. S.T., Y.W. and H.K. helped with the 2D-MoS$_2$ and 2D-WSe$_2$ MOCVD. E.M. and J.S. helped with the XPS measurement. A.P. helped with the wet transfer. A.V., M.H. and M.C.L. did the funding acquisition. A.V. and H.K. supervised the project. A.G. wrote the manuscript. All authors discussed the results and commented on the manuscript.

**Notes**

The authors declare no competing financial interests

ACKNOWLEDGEMENT

This work has been financially supported by German Federal Ministry of Education and Research (BMBF), in the scope of NEUROTEC II (no. 16ME0399) and NeuroSys (no. 03ZU1106AA) projects.

# Supporting Information

# Dry Transfer Based on PMMA and Thermal release Tape for Heterogeneous Integration of 2D-TMDC Layers

*Amir Ghiami,[†,*] Hleb Fiadziushkin,[†] Tianyishan Sun,[†] Songyao Tang,[†] Yibing Wang,[†] Eva Mayer,[‡] Jochen M. Schneider,[‡] Agata Piacentini,[§,α] Max C. Lemme,[§,α] Michael Heuken,[†,§] Holger Kalisch,[†] Andrei Vescan [†]*

[†] Compound Semiconductor Technology, RWTH Aachen University, 52074 Aachen, Germany
[‡] Materials Chemistry, RWTH Aachen University, 52074 Aachen, Germany
[§] Advanced Microelectronic Center Aachen (AMICA), AMO GmbH, Aachen, Germany
[α] Chair of Electronic Devices, RWTH Aachen University, 52074 Aachen, Germany
[§] AIXTRON SE, 52134 Herzogenrath, Germany
[*] Email: ghiami@cst-rwth-aachen.de

**Transfer onto prepatterned substrate**

To further demonstrate the efficacy of the developed transfer process, an MOCVD 2D-MoS$_2$ layer was successfully transferred onto a prepatterned Si-SiO$_2$ substrate with Au metal contacts of 50 nm height (see Figure S1). This can be utilized for the fabrication of bottom-contact FET[1] and memristors[2].

Figure S1 shows the prepatterned substrate before and after the 2D-MoS$_2$ transfer and PMMA removal, achieving a near-perfect transfer success rate, with no visible pinhole or defects under optical microscope. This demonstrates that our dry transfer method can be successfully applied not only to Si-SiO$_2$ substrates but also to prepatterned Si-SiO$_2$ substrate with metal contacts, without adhering to the 30 nm limit mentioned by Kwon et al.[3]

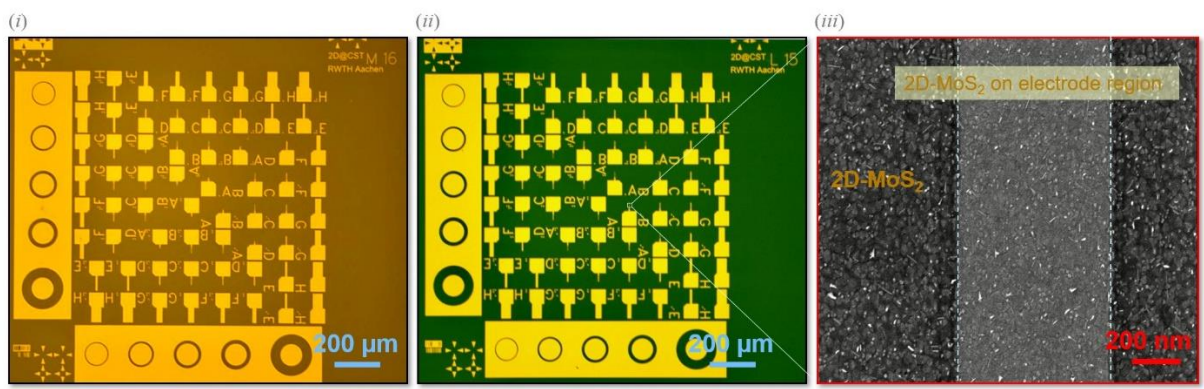

**Figure S1.** Dry transfer of the 2D-MoS$_2$ film onto the prepatterned substrate. *(i)* optical image of back electrodes fabricated on Si-SiO$_2$ substrate via photolithography and lift-off process, *(ii)* optical image of 2D-MoS$_2$ dry-transferred onto the prepatterned substrate (after PMMA removal), *(iii)* SEM image of the tip of an electrode with 2D-MoS$_2$ laying on top of it.

**XPS analyses of 2D-TMDC layers**

XPS analyses of as-deposited, wet-, and dry-transferred (after PMMA removal) 2D-MoS$_2$ and 2D-WSe$_2$ were performed. High resolution Mo$3d$ and S$2p$ core levels for 2D-MoS$_2$, and W$4f$ and Se$3d$ core levels for 2D-WSe$_2$ are shown in Figure S2 *(i)-(iv)*. For 2D-MoS$_2$, the Mo$3d$5/2, Mo$3d$3/2 and S$2s$ peaks are located at 229.2 *eV*, 232.4 *eV* and 226.3 *eV*, while the S$2p$3/2 and S$2p$1/2 peaks are located at 162.0 *eV* and 163.2 *eV*, respectively, in good agreement with the values reported in literature.[4,5]

The lineshapes verify the presence of pure MoS$_2$ in all as-deposited and transferred layers with neither oxide nor carbide species, which would have appeared at higher and lower binding energies relative to MoS$_2$ peaks, respectively. For 2D-WSe$_2$, W$4f$7/2, W$4f$5/2 and W$5p$3/2 peaks are located at 32.6 *eV*, 34.9 *eV* and 38.3 *eV*, and the Se$3d$5/2 and Se$3d$3/2 peaks are located at 54.9 *eV* and 55.8 *eV*, respectively, also in good agreement with the values reported in literature.[6,7] Similar to 2D-MoS$_2$, 2D-WSe$_2$ shows no oxidation peak, which would have appeared at higher binding energies relative to WSe$_2$ peaks. However, a peak located at lower binding energies relative to WSe$_2$ peaks can be detected, which is attributed to WSe$_{2-x}$C$_x$ species likely co-deposited during the MOCVD cool-down stage. This species is partially removed during the transfer, as suggested by its lower intensity in the transferred layers.

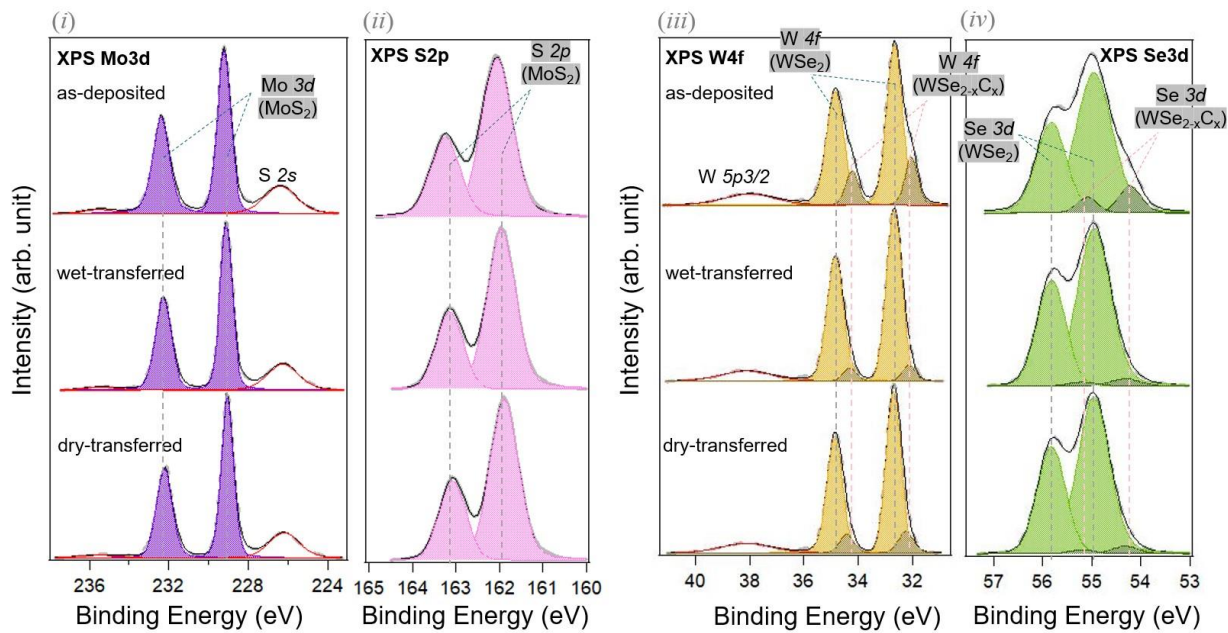

**Figure S2**. XPS analyses of as-deposited and transferred 2D-MoS$_2$ and 2D-WSe$_2$ layers. *(i)* Mo*3d*, *(ii)* S*2p*, and *(iii)* W*4f*, *(iv)* Se*3d*